\documentclass[prd,showpacs]{revtex4}
\usepackage{graphicx}
\usepackage{latexsym}
\usepackage[all]{xy}

\newcommand{\bea}{\begin{eqnarray}}
\newcommand{\eea}{\end{eqnarray}}
\newcommand{\be}{\begin{equation}}
\newcommand{\ee}{\end{equation}}

\begin{document}

\title{Dynamics of nonlocal cosmology}
\author{Tomi Koivisto}
\email{T.Koivisto@thphys.uni-heidelberg.de}
\affiliation{Institute for Theoretical Physics, University of Heidelberg, 69120 Heidelberg,Germany}

\begin{abstract}

Nonlocal quantum corrections to gravity have been recently proposed as a possible solution to the cosmological fine-tuning problems. 
We study the dynamics of a class of nonlocal actions defined by a function of the inverse d'Alembertian of the Ricci scalar.
Power-law and exponential functions are considered in detail, but we also show a method to reconstruct 
a nonlocal correction that generates a given background expansion. We find that even the simplest terms can, 
while involving only Planck scale constants, drive the late time acceleration without changing early cosmology. 
This leads to a sudden future singularity, which however may be avoided by regularizing the d'Alembertian. 
We also consider the Einstein frame versions of these models. 

\end{abstract}  

 
\maketitle

\section{Introduction}

The observed acceleration of the universe requires the presence of a tiny energy density that corresponds 
to a cosmological constant of the order $\Lambda \sim 10^{-120}$ in Planck units. This has motivated the consideration of 
dynamical alternatives to the cosmological constant \cite{Copeland:2006wr}. These alternatives usually introduce new fields
\cite{Uzan:2006mf}, by either adding quintessence in the matter sector or by modifying gravity. In the quintessence 
models the constant term is promoted to a potential function $\Lambda \rightarrow m_\varphi\varphi^2$ of the cosmon field $\varphi$ 
\cite{Wetterich:1987fm}, and in the modified gravity models one instead considers a curvature-dependent effective cosmological term $R + 
\Lambda \rightarrow R + \mu^{2(n+1)}/R^n$ \cite{Carroll:2003wy}. Typically one still requires an explanation for the tiny constants 
$m_\varphi,\mu \sim 10^{-33}$ eV within both of these approaches (in addition to the vanishing of the usual $\Lambda$). Recently a new, 
nonlocal variant was suggested \cite{Deser:2007jk} in the form $R + \Lambda \rightarrow R + Rf(\Box^{-1}R)$, where the function 
$f(\Box^{-1}R)$ of the inverse d'Alembertian acting on the curvature could involve constants of (roughly) the order one in Planck units. 
This seems a worthwhile starting point.

Quantum loop effects are well known to generate nonlocal corrections to the effective action of gravity \cite{Donoghue:1994dn}. 
Nonlocality arises naturally in string theory, and is considered as the most likely resolution of the black hole information paradox 
\cite{Giddings:2006sj}. Cosmological effects have been shown to arise from nonlocal corrections which were proposed as a cure to the 
unboundedness of the Euclidean gravity action \cite{Wetterich:1997bz}. The resulting curvature dependence of the 
effective Newton's constant may then easily pass the local gravity experiments, but could be constrained by 
its cosmological effects, especially on nucleosynthesis. String-inspired forms of nonperturbative gravity could also  
be asymptotically free and thereby describe bouncing cosmologies \cite{Biswas:2005qr,Biswas:2006bs}. The cosmological impact of the 
particular (logarithmic) form of quantum corrections expected at the leading order has been computed and a small possible 
effect on the expansion rate was found \cite{Espriu:2005qn,Cabrer:2007xm}. However, since it is a nonrenormalizable theory, 
long-distance modifications of gravity which could become significant at low curvature cannot be excluded, 
and in particular they could be responsible for the observed acceleration of the universe 
\cite{Barvinsky:2003kg,Hamber:2005dw,Deser:2007jk}. 

Let us note also that nonlocal cosmology has gained recently interest in the context of string field theory, with various  
applications to incorporate bounces, model dark energy and generate nongaussianities during inflation
\cite{Aref'eva:2004vw,Aref'eva:2006et,Aref'eva:2007uk,Aref'eva:2007yr,Calcagni:2005xc,Calcagni:2007ru,Barnaby:2007yb,Barnaby:2008fk}. 
For recent clarifications about the degrees of freedom in nonlocal field theories, see \cite{Barnaby:2007ve,Calcagni:2007ef}.
 
In this paper we focus on the cosmological consequences of the class of actions describing additions to 
general relativity as a function $f$ of the inverse d'Alembertian (i.e. covariant Laplacian) acting on the curvature (i.e. Ricci) scalar. The 
dynamics of the cosmological background expansion is studied with the main aim to find solutions naturally incorporating the present acceleration
of the universe. Then it is crucial to uncover in detail whether these models could feature a viable sequence of radiation dominated, 
matter dominated and accelerating eras. In particular, we then study the dynamics in two simple cases: when the function $f$ is 
of the 
power-law and of the exponential form. We are then able to find the conditions for the late acceleration to occur, and to determine the 
parameters of the model required for observationally consistent evolution of the universe. We also consider an alternative approach, where we 
decide on the background cosmological expansion history and reconstruct the form of the function $f$ to generate the given expansion. 
In general, such a procedure is difficult to perform analytically, and it may yield functions which are very complicated, unstable or 
singular. Therefore we mainly consider the approach of choosing a simple form of the function and studying its consequences, which can be 
then uniquely determined. In fact, we find that the evolution is qualitatively similar regardless of whether $f$ is exponential or a 
power-law.

In these models the acceleration of universe does not lead into a de Sitter stage, but instead ends in a singularity, where
the pressure diverges at a finite time, when the scale factor has expanded a finite amount. This is a so called future sudden singularity 
\cite{Barrow:2004xh}, which is different from the big rip singularities present in phantom dark energy models, described by 
a diverging scale factor at some finite moment of time. According to the classification of Ref.\cite{Nojiri:2005sx}, the sudden future 
singularity is labeled type II. We show that in the singularities may be avoided by regularizing the d'Alembertian operator. 

The plan of the paper is as follows. In Section \ref{sektio} we write down the basic equations, introducing the model 
and it's presentation as a biscalar-tensor theory in \ref{multi} and writing down the cosmological equations in \ref{cosmo}. We also 
discuss the Einstein
frame version of the theory in \ref{einstein}. In Section \ref{tosisektio} we then study cosmological dynamics
from various points of view. The initial conditions are conceptually important in these theories, and in this particular
class of models we can set the initial conditions at the radiation-dominated era according to the analytical solutions presented in subsection
\ref{radi}. We find also analytically the conditions for the matter-dominated era to end. These are derived in subsection \ref{mati} for the 
power-law and the exponential model. Numerical computations show the details of the evolution and reveal the generality of the singularity.
The approach to singularity and its avoidance is studied analytically in a particular case in subsection \ref{dari}. An alternative
route to finding solutions is given in subsection \ref{reco} as a reconstruction method and it is applied in a discussion
of scaling solutions in the special case of exponential $f$. In Section \ref{conclu} we present our conclusions.  


\section{The nonlocal model}
\label{sektio}

In this Section we specify our nonlocal action and rewrite it as a multi-scalar-tensor theory \cite{Damour:1992we} with extra 
kinetic couplings. The cosmological equations are then written as generalizations of the standard Friedmann and continuity
equations, and this is again recasted into more useful form by a change of variables. Finally, the corresponding 
system is considered in the Einstein frame.


\subsection{Multi-scalar-tensor theory}
\label{multi}

We consider the following simple example of a nonlocal action \cite{Deser:2007jk,Nojiri:2007uq}
\be
S  =  \int d^n x \sqrt{-g}\left[ R\left(1+f(\Box^{-1}R)\right) + 2\kappa^2\mathcal{L}_m \right]. \label{eka}  
\ee
Introducing the Lagrange multiplier $\xi$ to rename the inverse d'Alembertian as $\phi$, we may rewrite the action in the
form 
\be
S  =  \int d^n x \sqrt{-g}\left[R\left(1+f(\phi)\right) + \xi(\Box\phi -R) + 2\kappa^2\mathcal{L}_m\right]. 
\ee
More general action involving $m$ different powers $\Box$ acting on $R$ can be recasted into a multi-scalar-tensor theory by
introducing $2m$ scalar fields \cite{Nojiri:2007uq}. Here we restrict to the case where only the first power $\Box^{-1}$ appears in 
the action.
Integrating partially and neglecting the boundary terms, as usual, we can then eliminate the double derivative, 
\be
S  =  \int d^n x \sqrt{-g}\left[R\left(1+f(\phi)-\xi\right) - \nabla_\alpha\xi\nabla^\alpha\phi + 2\kappa^2\mathcal{L}_m\right]. \label{s3}
\ee
This is a peculiar kind of double-scalar-tensor theory, where massless, nonminimally coupled scalars have a kinetic-type interaction resembling 
of the nonlinear sigma model. From the form (\ref{s3}) it is clear that one may not eliminate the kinetic coupling by a field redefinition. 
However, it turns out to be convenient to make yet one more manipulation of the action. If we use, instead of the field $\xi$, the field 
\be
\Psi(\phi,\xi) \equiv f(\phi) - \xi
\ee
as our dynamical variable, it becomes transparent that we may consider a double-scalar-tensor theory, with only one of the scalar degrees of 
freedom is nonminimally coupled to gravity. The pair of the scalar fields still has the unusual kinetic-type coupling\footnote{By 
the rescaling $\phi \rightarrow 
\int\sqrt{2f'(\phi)}d\phi$ one could yet turn the kinetic term of $\phi$ into the canonical
form, but this would not be helpful in the following.},
\be
S =  \int d^n x \sqrt{-g}\left[(1+\Psi)R - 
f'(\phi)(\partial\phi)^2 + \nabla_\alpha\Psi\nabla^\alpha\phi + 2\kappa^2\mathcal{L}_m\right].
\ee
We get the field equations
\bea
G_{\mu\nu} & = & \kappa^2 T_{\mu\nu} - \frac{1}{2}\nabla_\alpha\xi\nabla^\alpha\phi g_{\mu\nu} + \nabla_{(\mu}\phi\nabla_{\nu)}\xi - 
            \left(g_{\mu\nu}\Box-\nabla_\mu\nabla_\nu-G_{\mu\nu}\right)(f(\phi)-\xi) \\ 
           & = & \kappa^2 T_{\mu\nu} + T^\phi_{\mu\nu} + T^\Psi_{\mu\nu} + T^X_{\mu\nu},  \label{efe}
\eea
where the effective energy momentum tensors for the two fields and for their coupling read as
\bea
T^\phi_{\mu\nu} & = & f'(\phi)\left[-\frac{1}{2}(\partial\phi)^2 g_{\mu\nu} + \nabla_\mu\phi\nabla_\nu\phi\right],  \label{set_phi} \\
T^\Psi_{\mu\nu} & = & -\left(G_{\mu\nu} + g_{\mu\nu}\Box-\nabla_\mu\nabla_\nu \right)\Psi,                          \label{set_psi} \\ 
T^X_{\mu\nu} & = & \frac{1}{2}\nabla_\alpha\Psi\nabla^\alpha\phi g_{\mu\nu} - \nabla_{(\mu}\phi\nabla_{\nu)}\Psi.   \label{set_x}
\eea 
The first piece reduces to canonical kinetic terms, when $f=\phi$. 
Because of the additional coupling to the other field, $f'(\phi)>0$ does not now necessarily imply that the field is a ghost. 
The second part corresponds to the usual nonminimal coupling of a standard 
Brans-Dicke field $\Psi$, and is also well-known to be consistent contribution to 
field equations both at the quantum and classical level given $\Psi>-1$. The last piece is linear in the 
derivatives of both the scalar fields, which may not be an unproblematical feature. 

The equations of motion are
\bea \label{fields}
\Box\Psi & = & f''(\phi)(\partial\phi)^2 + 2f'(\phi)R, \\
\Box \phi & = & R,
\eea  
where the Ricci curvature could be solved from the trace of field equations,
\be \label{trace}
(1-\frac{n}{2})(1+\Psi) R = \kappa^2 T - (\frac{n}{2}-1)\left(f'(\phi)(\partial\phi)^2 - \nabla_\alpha\Psi\nabla^\alpha\phi\right) - (n-1)\Box \Psi.
\ee
Since minimally coupled, the matter tensor is conserved, $\nabla^\mu T^m_{\mu\nu}$, which may be used to check the system of 
equations (\ref{efe}-\ref{trace}) as it is not an independent condition \cite{Koivisto:2005yk}. 

Let us consider the case $f = f_0 R/\Box$. From the equations of motions (\ref{fields}) one sees that $\Box \phi$ is now proportional to the 
$\Box\Psi$. It is thus possible to eliminate one degree of freedom. This is similar to the so called nonlinear gravity theories 
defined by a function of $R$: in such a class of theories there is an extra scalar degree of freedom, which vanishes only in the 
unique case of the Einstein-Hilbert action that is linear in $R$. Here we then noticed that multiplying the curvature term by a 
function of $\Box R$ is equivalent to introducing two additional scalar fields, and the unique case of a linear function again 
allows one to integrate out one scalar degree of freedom.

One could also consider generalizing the D'Alembertian operator $\Box$ with regulating terms like 
$\Box - \epsilon R + (\gamma/\kappa^2) R_{\mu\nu\rho\sigma}R^{\mu\nu\rho\sigma} \dots$, where $\epsilon$ and $\gamma$ are 
dimensionless numbers \cite{Wetterich:1997bz}. If we make the substitution $\Box^{-1} \rightarrow (\Box - h(R))^{-1}$ 
in the original action (\ref{eka}), by similar steps as before we end up with a generalized Lagrangian, 
\be  \label{h_lag}
S = \int d^n x \sqrt{-g}\left[R\left(1+f(\phi)-\xi - h'(\chi)\xi\phi\right) + 
(h'(\chi)\chi-h(\chi))\xi\phi - \nabla_\alpha\xi\nabla^\alpha\phi + 2\kappa^2\mathcal{L}_m\right]. 
\ee 
If one considers a linear function $h(\chi)$, it is possible to eliminate the degree of freedom $\chi$. Even with a general $h(\chi)$, in
the case of linear $f(\phi)$, one may eliminate one degree of freedom since $\xi$ can be related to $\phi$. Thus, with linear $f$ and linear $h$,
Eq.(\ref{h_lag}) becomes an action for a single scalar nonminimally coupled to gravity.  
We return to this interesting special case in \ref{dari}.

\subsection{Cosmological equations} \label{cosmo}

Having now the general equations at hand, we may see how they become in cosmology.
In a flat Friedmann-Robertson-Walker (FRW) background,
\be
ds^2 = -dt^2 + a^2(t)d\mathbf{x}^2
\ee
we get the Friedmann equations 
\be
3H^2(1+\Psi)=\kappa^2\rho_m + \frac{1}{2}\left(f'(\phi)\dot{\phi}^2 - \dot{\Psi}\dot{\phi}\right) - 3H\dot{\Psi}
\ee
\be
-(2\dot{H}+3H^2)(1+\Psi) = \kappa^2p_m + \frac{1}{2}\left(f'(\phi)\dot{\phi}^2 - \dot{\Psi}\dot{\phi}\right) + \ddot{\Psi} + 2H\dot{\Psi}
\ee
The equations of motion for the fields are then
\be
\ddot{\phi}+3H\dot{\phi}=-6(\dot{H}+2H^2),
\ee
\be
\ddot{\Psi}+3H\dot{\Psi} = f''(\phi)\dot{\phi}^2 -12f'(\phi)(\dot{H}+2H^2).
\ee
Introduce the dimensionless variable 
\be
X \equiv \frac{\dot{\phi}}{2H} = \frac{\phi^*}{2} 
\ee
together with the e-fold derivative $y^* \equiv dy/d\log{a}$. 
We define the energy density fraction of matter as usual,
\be 
\Omega_m \equiv \frac{8\pi\rho_m}{3 H^2}, 
\ee
We then have a complete set of dimensionless variables. The dynamical system may then be written as  
\bea \label{om_evol}
\Omega_m^* & = & 3(w_m-w_{eff})\Omega_m, \\ \label{psi_evol}
\Psi^{**} & = & \frac{3}{2}(w_{eff}-1)\Psi^* - 6f'(\phi)(1-3w_{eff}) + 4f''(\phi)X^2, \\ \label{x_evol}
X^* & = & \frac{3}{2}\left[(w_{eff}-1)X  + 3w_{eff}-1\right].
\eea
where
\be
w_{eff} = \frac{3w_m\Omega_m  + 2f'(\phi)(X^2-3) + 4f''(\phi)X^2 - (1+X)\Psi^*}{3(1+\Psi-6f'(\phi))}.
\ee
The first equation simply expresses the usual matter conservation law, and the two following equations are the conservation laws for
our two scalar fields. 

The fixed points of the system, defined by $\Omega_m^*=\Psi^*=X^*=0$ are as follows.
\begin{itemize}
\item It is possible to construct scaling solutions with $w_{eff}=w_m$ and with arbitrary $\Omega_m$ if the coupling is of the form 
      $f(\phi) \sim e^{\alpha\phi}$, where $\alpha=3(1-w_m)^2/(2-6w_m)$. However, this is a very special case with not only the form
      but the parameters of the coupling fixed. Therefore we do not consider this possibility further.
\item In the case of effectively complete radiation domination, $w_{eff}=1/3$, we automatically have a fixed point, since with $X=0$ the equations
      are identically satisfied. The matter density and the field $\Psi$ are then related by the $\Omega_m=1+\Psi$. In the very early universe, 
      when we have exactly $w_{eff}=w_m=1/3$, the correct initial condition is $\Psi=0$. The evolution begins when $w_{m}$ drops below $1/3$, as we discuss 
      in the following subsection.  
\item If $f'=f''=0$ the equations are again identically satisfied with arbitrary $\Psi$, and now $X=3(1-w_{eff})^2/(2-6w_{eff})$ and $1+\Psi=\Omega_m$. 
      The most relevant case is of course $\Psi=0$, which corresponds to standard cosmology. The field $\phi$ keeps growing, as we see in the following,
      but it depends on the form of the coupling whether this means that the effect of nonlocal gravity is becoming significant.
\end{itemize}
In fact we note that the information about the fixed points in this system throws light only on some special corners of the phase space, and for more 
satisfactory understanding of the nonlocal dynamics we now turn into other considerations. 
The system of alternative variables normalized by $f(\phi)$, say $\Omega_m/f(\phi)$, would feature additional fixed points 
in the special case of exponential potential \cite{Jhingan:2008ym}. We will also find and generalize these same solutions in 
subsection (\ref{reco}).

\subsection{The Einstein frame}
\label{einstein}

Though we will work mainly in the usual Jordan frame, we will briefly consider the Einstein frame version of these models in this
subsection. Perform the Weyl rescaling \cite{Wands:1993uu,Faraoni:2006fx}
\be \label{weyl}
\tilde{g}_{\mu\nu} \equiv (1+\Psi)^{\frac{2}{n-2}}g_{\mu\nu} \equiv e^{\frac{2}{n-2}\varphi}g_{\mu\nu}. 
\ee
The action becomes
\be
\label{se}
S  =  \int d^n x \sqrt{-\tilde{g}}\left[\tilde{R} - \frac{n-1}{n-2}(\tilde{\nabla}\varphi)^2 - 
e^{-\varphi} f'(\phi)(\tilde{\nabla}\phi)^2 + \tilde{\nabla}_\alpha\varphi\tilde{\nabla}^\alpha\phi  + 
2\kappa^2e^{-2\varphi} \mathcal{L}_m(\tilde{g}e^{-\varphi}). \right]
\ee
Now $\Psi$ has also acquired a kinetic term. The curvature coupling has been removed in this frame, but the matter sector has now become nonminimally coupled.
Since there remains the derivative coupling term, things have not simplified so much as in the Einstein frame of the Brans-Dicke class of theories. 
However, the metric rescaling provides also here a useful alternative viewpoint to the system.
The field equation is
\be
\tilde{G}_{\mu\nu} =
\tilde{T}_{\mu\nu} +
\tilde{T}^\phi_{\mu\nu} + 
\tilde{T}^\varphi_{\mu\nu} + 
\tilde{T}^X_{\mu\nu}, 
\ee
where we have again decomposed the left hand side to four contributions,
\bea
\tilde{T}^\phi_{\mu\nu} & = & {f'(\phi)}e^{-\varphi}\left[-\frac{1}{2}(\tilde{\nabla}\phi)^2 \tilde{g}_{\mu\nu}
+ \tilde{\nabla}_\mu\phi\tilde{\nabla}_\nu\phi\right], \\
\tilde{T}^\Psi_{\mu\nu} & = & \left(1+\frac{1}{n-2}\right)\left[-\frac{1}{2}(\tilde{\nabla}\varphi)^2 
\tilde{g}_{\mu\nu}
+ \tilde{\nabla}_\mu\varphi\tilde{\nabla}_\nu\varphi\right], \\
T^X_{\mu\nu} & = & \left[\frac{1}{2}\tilde{\nabla}_\alpha\varphi\tilde{\nabla}^\alpha\phi \tilde{g}_{\mu\nu} - 
\tilde{\nabla}_{(\mu}\phi\tilde{\nabla}_{\nu)}\varphi\right].
\eea
Note that none of the four energy momentum tensors is conserved. Matter, as well as the field $\phi$, is coupled nonminimally to the field 
$\varphi$. The equation of motion for this field may be written as 
\be
\left(\frac{e^\varphi}{2f'(\phi)}-\frac{2(n-1)}{(n-2)}\right)\tilde{\Box}\varphi =
\left(\frac{f''(\phi)}{2f'(\phi)} + 
e^{-\varphi}\right)(\tilde{\nabla}\phi)^2 - \tilde{\nabla}_\alpha\varphi\tilde{\nabla}^\alpha\phi - 
\frac{\kappa^2\tilde{T}}{(\frac{n}{2}-1)e^{2\varphi/(n-2)}}.
\ee
This could be derived either by varying the action (\ref{se}) or by conformally transforming the equation (\ref{fields}) and using 
\be
\tilde{R} = \frac{1}{1-\frac{n}{2}}\tilde{T} + \frac{n-1}{n-2}(\tilde{\nabla}\varphi)^2
 - e^{-\varphi}f'(\phi)(\tilde{\nabla}\phi)^2 - e^{-\varphi}\tilde{\nabla}_{(\mu}\phi\tilde{\nabla}_{\nu)}\varphi .
\ee
Then, to find the equation of motion for the field $\phi$ in the Einstein frame, it is most convenient to note from Eqs.(\ref{fields})
that $\tilde{\Box}\phi = \frac{1}{2}(\tilde{\Box}\Psi-f''(\phi)(\tilde{\nabla}\phi)^2)/f'(\phi)$. 

The Einstein frame cosmology can then be obtained from the Jordan frame solutions by carefully taking into account the conformal dimension of each 
quantity. From the scaling (\ref{weyl}) we see that the scale factor, being a squareroot of a component of the metric, has conformal dimension
half, $\tilde{a}=e^{\varphi/2}a$. The invariance of the line-element then requires $d\tilde{t}=e^{\varphi/2}dt$. From this already
follows that the Hubble rate in the Einstein frame is
\be
\tilde{H} = e^{-\varphi/2}(H+\frac{1}{2}\dot{\varphi}),
\ee
and that its derivative may be written as  
\be 
\frac{1}{\tilde{H}^2}\frac{d\tilde{H}}{d\tilde{t}} = 
\frac{1}{1+2\varphi^*}\left[\frac{\varphi^{**}}{2+\varphi^*}+\left(\frac{H^*}{H}-\frac{1}{2}\varphi^*\right)\right]. 
\ee
This dimensionless quantity gives directly the effective equation of state in the Einstein frame. As a manifestation
of the matter nonconservation, we have $\tilde{\rho}_m = e^{-2\varphi}\rho_m$. It follows that
\be
\tilde{\Omega}_m = \frac{\Omega_m}{e^{\varphi}(1+\frac{1}{2}\varphi^*)^2}
\ee
gives the relative amount of matter.

\section{Cosmological dynamics}
\label{tosisektio}

\begin{figure}
\begin{center}
\includegraphics[width=0.45\textwidth]{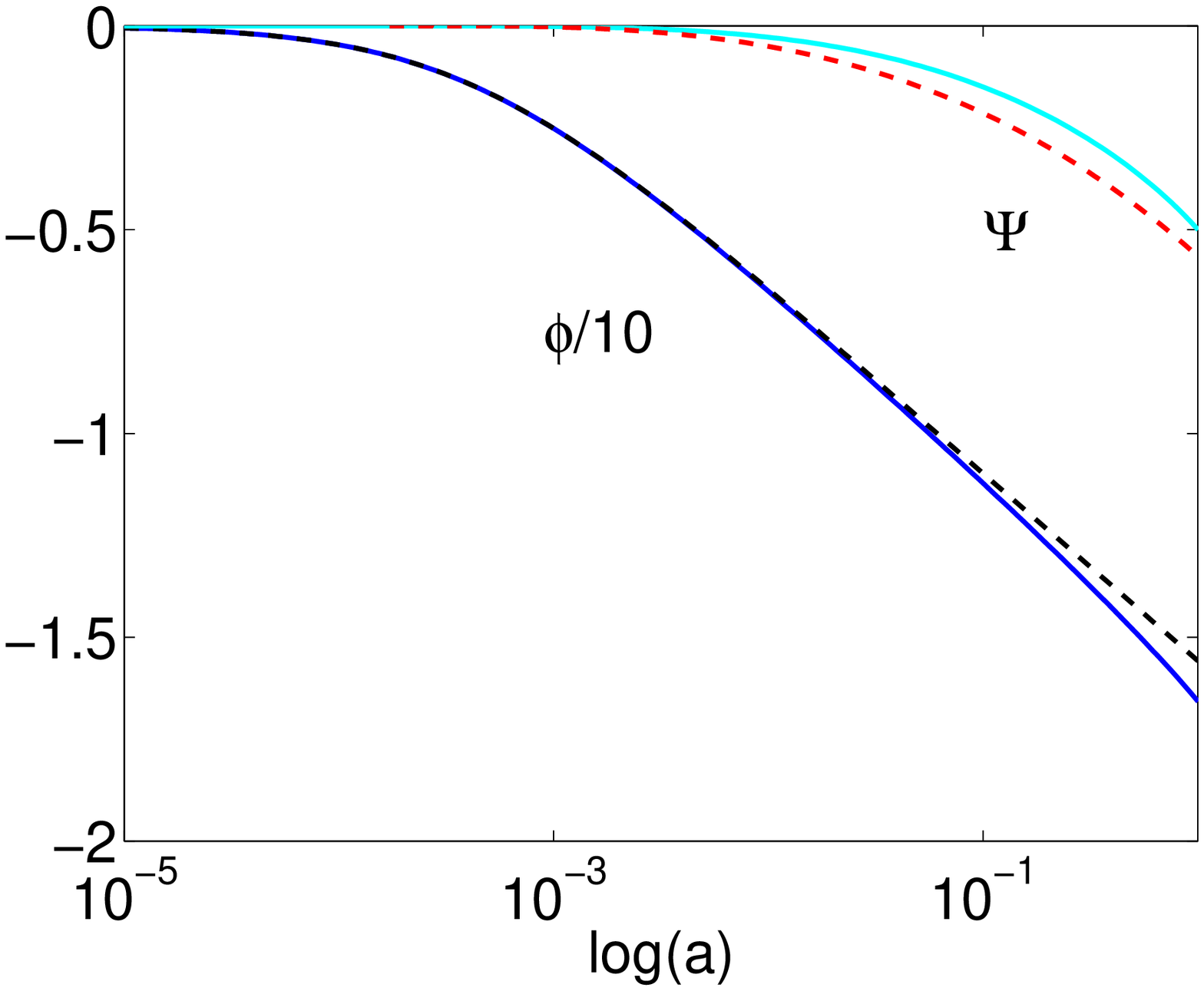}
\includegraphics[width=0.45\textwidth]{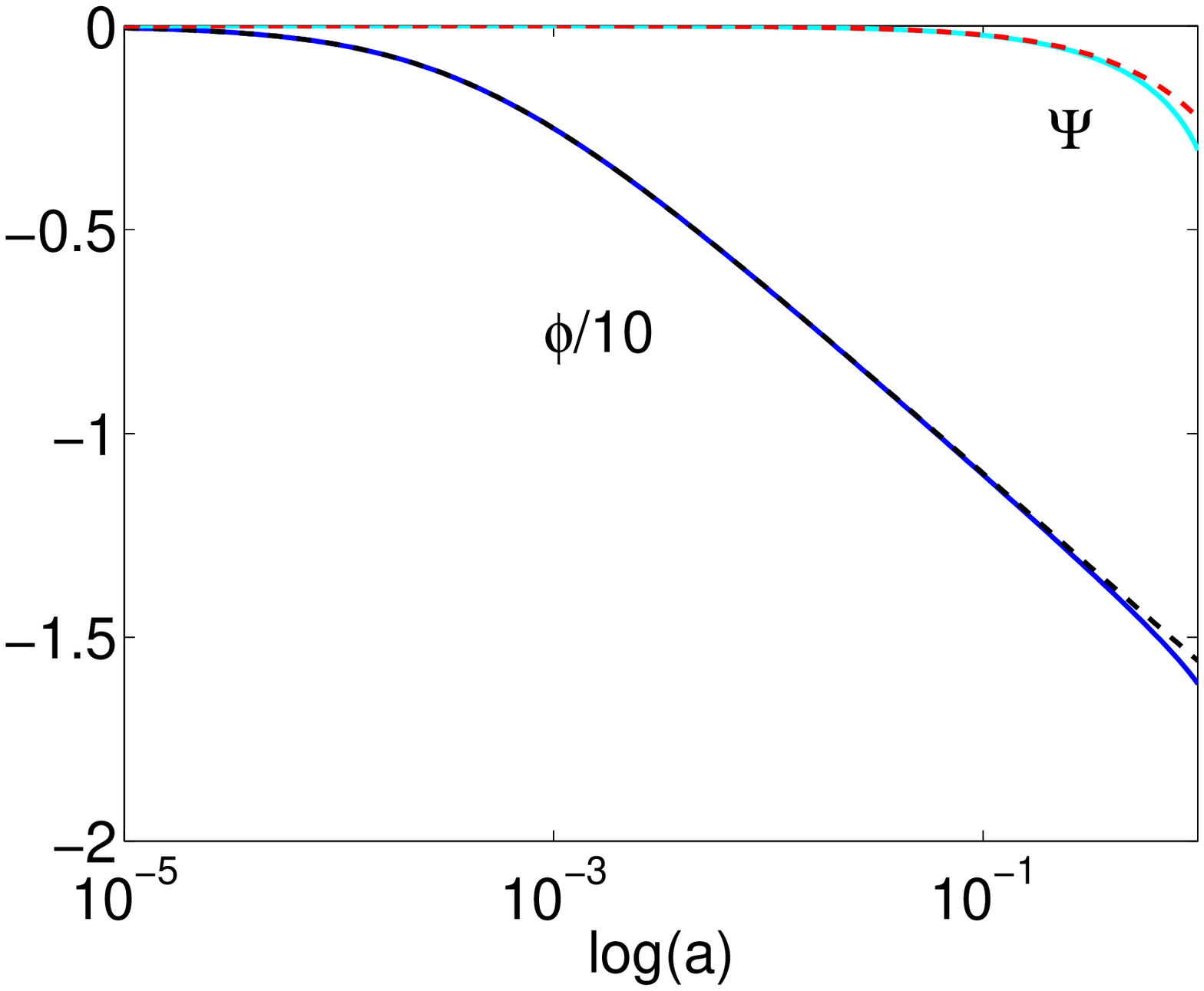}
\caption{\label{appro} Evolution of the scalar fields and comparison with the analytic approximations.
The left panel is the power-law model with $n=3$ (row 3 in Table \ref{tab}) and the right panel is the exponential
model with $\alpha=-0.5$.
The lower solid (blue) 
line is the numerical solution for $\phi$ (divided by ten to fit into figure), and the dashed (black) line is the approximation (\ref{ini1}). The 
upper solid (cyan) line is the numerical solution for $\Psi$, and the dashed (red) line the approximation (\ref{appro2}). We note that even 
after the universe has begun to accelerate, the matter-dominated solutions for the fields remain reasonable approximations.
}
\end{center}
\end{figure}

\begin{figure}
\begin{center}
\includegraphics[width=0.45\textwidth]{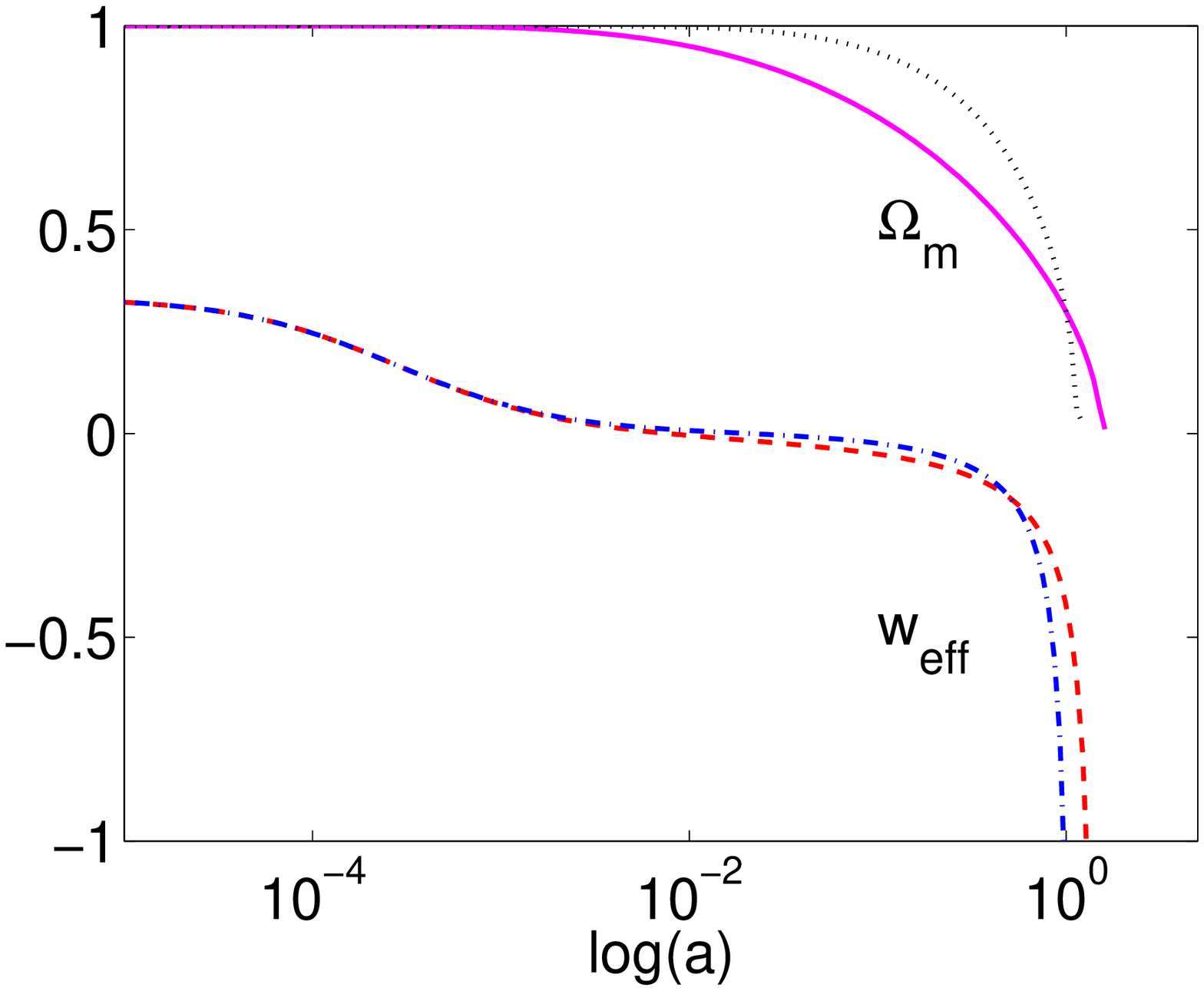}
\includegraphics[width=0.45\textwidth]{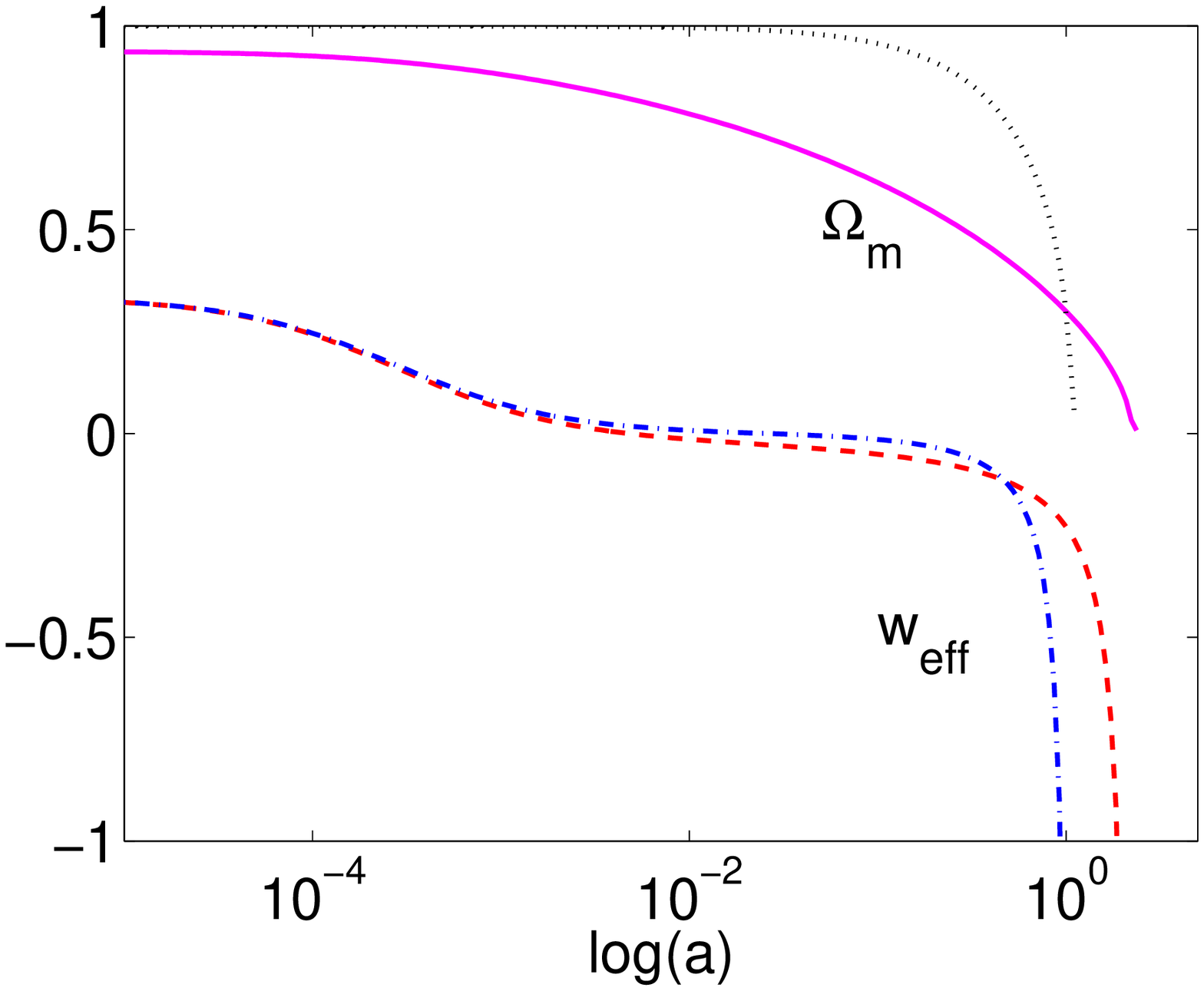}
\caption{\label{eosn} 
The cosmological evolution leading to the acceleration and singularity in various cases.
\newline
Left panel: 
The power-law model with $n=3$ and $n=6$ (rows 3 and 6 in Table \ref{tab}). 
The solid (magenta) line is $\Omega_m$, and the dashed (red) line is the $w_{eff}$ for $n=3$.
The dotted (black) line is $\Omega_m$, and the dash-dotted (blue) line is the $w_{eff}$ for $n=6$.
The onset of acceleration leads to a singularity in the effective equation of state. The evolution
is steeper for larger $n$. With smaller $n$, the relative amount of matter $\Omega_m$, begins to decrease earlier
and $w_{eff}$ drops later to minus infinity.
\newline
Right panel: 
The exponential model with $\alpha=-0.1$ and $\alpha=-0.5$. 
The solid (magenta) line is $\Omega_m$, and the dashed (red) line is the $w_{eff}$ for $\alpha=0.1$.
The dotted (black) line is $\Omega_m$, and the dash-dotted (blue) line is the $w_{eff}$ for $\alpha=0.5$.
The onset of acceleration leads to a singularity in the effective equation of state. The evolution
gets steeper for more negative $\alpha$. To set $\Omega_m=0.3$ today fixes $f_0= -0.296$ for  $\alpha=-0.1$
and $f_0=-0.140$ for $\alpha=-0.5$.}
\end{center}
\end{figure}

In this Section we will study cosmology at four different epochs: radiation dominated, matter dominated, accelerating, and a general
scaling. We can then find different approximations which are helpful in analytical considerations in these four cases.  
Numerical solutions are given in table \ref{tab} and the figures \ref{appro} and \ref{eosn}. 

\subsection{Radiation domination: Initial conditions}
\label{radi}

Consider the radiation dominated epoch. We assume that the nonlocal corrections are negligible. By definitions, we may always write 
\be \label{ricci} 
R=3(1-3w_{eff})H^2.
\ee
Assuming standard gravity with matter and radiation as sources, we have
\be \label{expansion} 
3w_{eff} = \frac{1}{1+ra}, \quad \left(\frac{H}{H_0}\right)^2 = \frac{\Omega^0_r}{a^4}(1+ra), 
\ee
where the fraction of nowadays relative matter density, $\Omega^0_m = 3\rho^0_m/(8\pi G H_0^2)$ and nowadays relative 
radiation density $\Omega^0_r=3\rho^0_r/(8\pi G H_0^2)$ is denoted $r \equiv \Omega^0_m/\Omega^0_r$. When all matter is 
relativistic, $R$ vanishes identically. Thus $\phi$ should also vanish. When part of the matter becomes nonrelativistic,
$R$ begins to evolve (even if the universe is dominated by radiation). The scalar field $\phi$ is given as the solution to 
the equation $\phi = \Box R$. Explicitly, we get
\be \label{di}
\phi(t) = -\int_{t_i}^t \frac{dt'}{a^3(t')}\int_{t_i}^ta^3(t'')R(t'')dt'' + 
\dot{\phi}_i\int_{t_i}^t\left(\frac{a_i}{a(t')}\right)^3dt' + \phi_i.
\ee
Note that this does not depend on the form of $f(\phi)$, since the scalar $\phi$ is just a way of reexpressing $R/\Box$. 
We consider exclusively the retarded solutions to respect causality.
In our models, we then set $\phi_i=\dot{\phi}_i=0$ at an initial, completely radiation dominated time $t_i$. When $R$ vanishes, $\phi=\Box R$ 
should also together with its derivatives. Using the expansion (\ref{expansion}), we can then reduce the double integral (\ref{di}) into the 
form
\be 
\phi(a) = -3r\int_{a_i}^a\frac{da'}{{a'}^2\sqrt{1+ra'}}\int_{a_i}^{a'}\frac{a''da''}{\sqrt{1+ra''}}
\ee
which one may perform analytically, yielding
\be \label{ini1}
\phi(a) = \frac{4(a-a_i)}{raa_i}-2\log{\frac{a}{a_i}} + 
(ra_i-2)\sqrt{1+ra_i}\left(\log\frac{(1+\sqrt{1+ra})(1-\sqrt{1+ra_i})}{(1-\sqrt{1+ra})(1+\sqrt{1+ra_i})}  
- \frac{2\sqrt{1+ra}}{ra} + \frac{2\sqrt{1+ra_i}}{ra_i}\right). 
\ee
Obviously, the solution depends on the moment $a_i$ at which we assume the matter in the universe to become nonrelativistic. 
One then obtains the derivative of the field,
\be \label{ini2}
X(a) = \left(2 + (ra_i-2)\frac{\sqrt{1+ra_i}}{\sqrt{1+ra_i}}\right)\frac{1}{ra} - 1. 
\ee 
Using the solution for $\phi$, the other scalar field $\xi$ may then be computed as
\be \label{di2}
\xi(t) = \int_{t_i}^t \frac{dt'}{a^3(t')}\int_{t_i}^ta^3(t'')R(t'')f(\phi(t''))dt'' + 
\dot{\xi}_i\int_{t_i}^t\left(\frac{a_i}{a(t')}\right)^3dt' + \xi_i.
\ee
The value of $\xi$ thus depends on the form of $f(\phi)$, and has to be therefore considered separately for each form of the nonlocal 
term $f$. Analytic solutions may be difficult to find, unlike for $\phi$. Once $\xi$ is then solved, 
it is however straightforward to obtain $\Psi=f-\xi$, and its time first derivative. These then completely fix the initial conditions for our 
system. 

In principle, due to the nonlocal property, the initial conditions could carry information from inflation or even from a pre-Planck era, if 
there effectively $R \neq 0$. Here we just assume that $\phi$ begins to evolve from zero when matter becomes nonrelativistic. Furthermore,
we have assumed above that this happens instantaneously at a given epoch $a=a_i$ (in our numerical examples this is $a_i=e^{-23}$). 
In reality each particle species begins to contribute to the trace $T_m$ when the temperature of the 
universe has dropped below the mass of the particle. Thus a more precise calculation should consider $r$ replaced by $r_*(a)$ in 
Eq.(\ref{expansion}) and 
elsewhere, $r_*(a)$ being a function which is zero as $a\rightarrow 0$, but is ''kicked'' to a slightly larger value each time 
some particle becomes nonrelativistic, and thus only eventually reaches $r_*(a) \rightarrow r$. Here we have approximated this a function
as the step function $r_*(a) = r\theta(a-a_i)$, since the thermodynamic details of the very early universe seem not to be important for our 
conclusions. This is because the solution (\ref{ini2}) is an attractor. One may note this easily by inserting $X+\delta X$ to 
the Eq.(\ref{x_evol}), which then yields
\be
(\delta X)^* = -\frac{1}{2}\left(3-\frac{1}{1+ar}\right)\delta X.
\ee
Thus the evolution of $\phi$ tends to a similar track with the solution (\ref{ini1}). A different initial condition will only set a different 
normalization for the field. However, since values of the field in Eq.(\ref{ini1}) will be of the order of $\phi \sim 10^{-7}$ in the early 
universe, these differences are negligible when the correction terms become important. Note however, that in the case where the nonlocal 
corrections would be large at high $R$, one should be careful with the early radiation history.

\subsection{Matter domination: Conditions for growth of nonlocalities}
\label{mati}

At late enough matter era, when $a \gg a_i$, the field (\ref{ini1}) is approximated by
\be \label{mlimit}
\phi(a) = -2\left(\log{\frac{ra}{4}} + 1 \right). 
\ee
This limit also confirms our claim that at late enough era the dependence on the initial conditions is negligible, since the $a_i$ is canceled 
out from the limit (\ref{mlimit}). The field derivative during matter domination follows immediately,
\be \label{mlimitx}
X(a) = - 1.
\ee
This means that during an e-fold $\Box^{-1}R$ drops by $-2$. To analyze the evolution of the other field, we must then specify
the function $f$. We consider two cases: an exponential and a power-law form.

\subsubsection{Exponential model} 
\begin{figure}
\begin{center}
\includegraphics[width=0.45\textwidth]{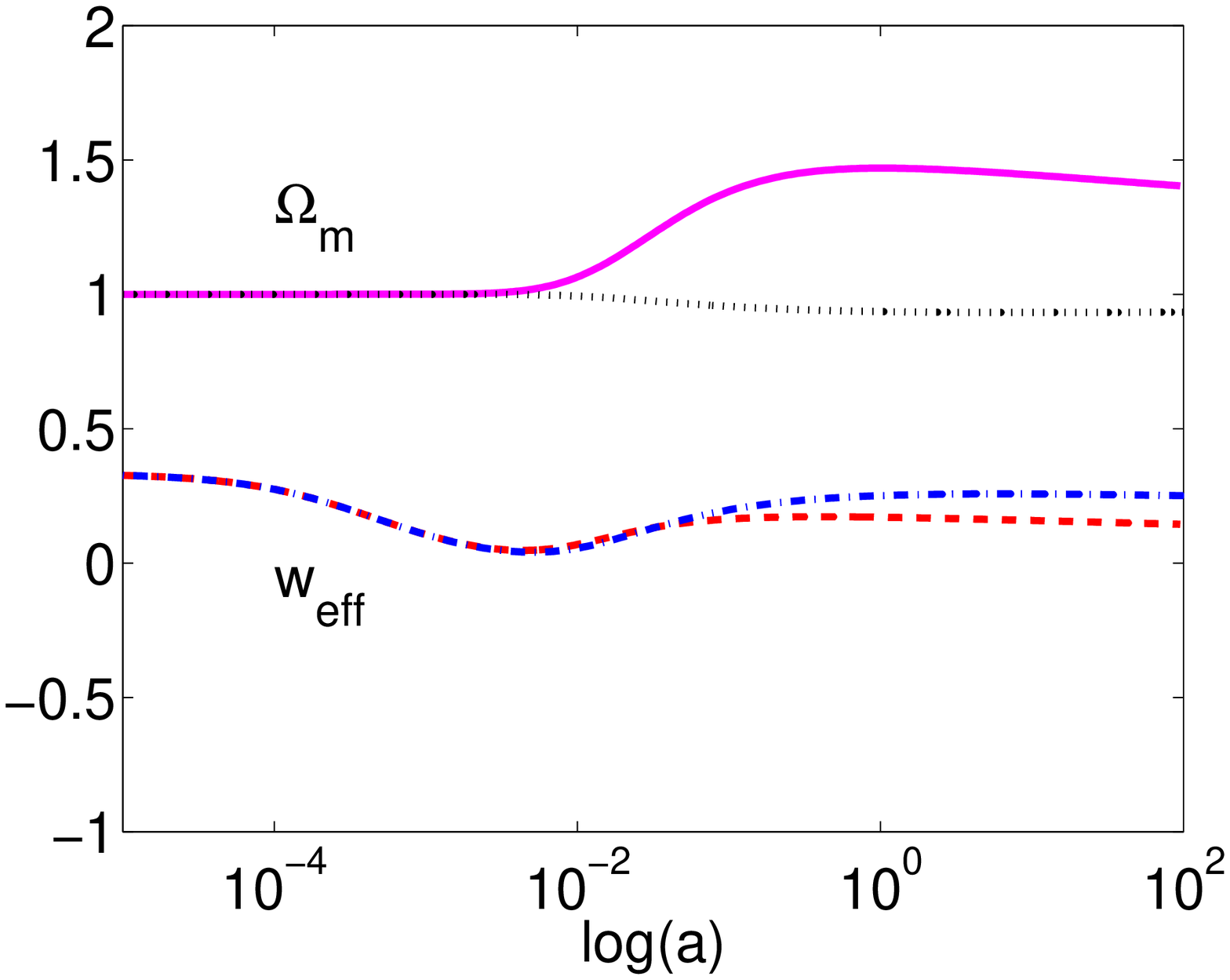}
\includegraphics[width=0.45\textwidth]{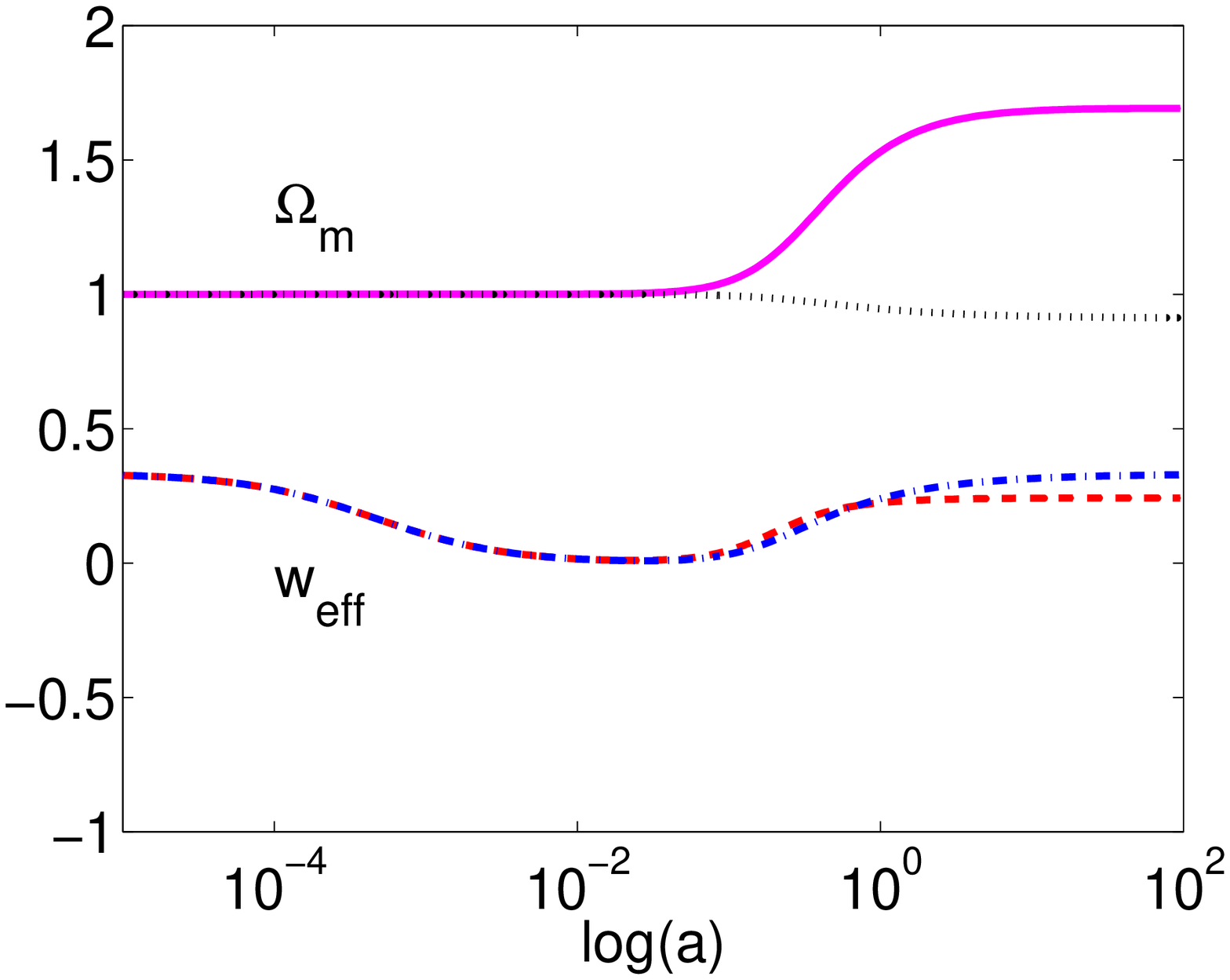}
\caption{\label{eos_no} 
The cosmological evolution leading to (nonaccelerating) effects from nonlocal corrections.
\newline
Left panel: 
The power-law model with $\alpha=6.0$ and $f_0=10^{-6}$.  
The solid (magenta) line is $\Omega_m/(1+\Psi)$, and the dashed (red) line is the $w_{eff}$.
One notes that the effect of $f$ slowly decays and the universe asymptotically
returns to the standard Einstein-de Sitter stage.
For curiosity, we plot also the evolution in the Einstein frame. 
The dotted (black) line is $\tilde{\Omega}_m$, and the dash-dotted (blue) line is the $\tilde{w}_{eff}$, and
These are understood as functions of the logarithm of the Einstein frame scale factor $\tilde{a}$.
\newline
Right panel: 
The exponential model with $\alpha=-1.0$ and $\hat{f}_0=10^{-6}$.  
The solid (magenta) line is $\Omega_m/(1+\Psi)$, and the dashed (red) line is the $w_{eff}$.
This is a scaling solution where we may consistently have $\Omega_m/(1+\Psi)>1$ because there can be negative 
effective energy sources. Here $w_{eff}$ settles to the value given by (\ref{scaling}). 
The evolution is plotted also in the Einstein frame.
The dotted (black) line is $\tilde{\Omega}_m$, and the dash-dotted (blue) line is the $\tilde{w}_{eff}$, and
these are understood as functions of the logarithm of the Einstein frame scale factor $\tilde{a}$. 
In both frames we find asymptotic expansion with a nonzero constant effective equation state for the exponential model.}
\end{center}
\end{figure}

Consider the exponential case, $f(\phi) = f_0e^{\alpha\phi}$. 
One may note that the pure exponentials models are ruled out because they feature a constant shift of the effective Newton's constant by
$f_0$, since the constant $(1+f_0)/(2\kappa^2)$ then multiplies the standard Einstein-Hilbert term. This however may be avoided by simple 
reformulations of the model, and the exponential case is still an interesting example.
We can now write 
\be
f(a) = \hat{f}_0a^{-2\alpha}, \quad \hat{f}_0 \equiv \left(\frac{4}{re}\right)^{2\alpha}f_0. 
\ee
One can then easily see that the leading contribution in the integral (\ref{di2}) is growing with $a$ proportional to $\xi \sim a^{-2\alpha}$ 
only when $\alpha<0$ The solution is then
\be
\Psi = \frac{6-4\alpha}{3-4\alpha}\hat{f}_0a^{-2\alpha}. 
\ee
This solution satisfies the evolution Eq.(\ref{psi_evol}) in the limit $w_{eff}=0$. 
Since the coupling is growing, the correction term will inevitably become significant at some point. However, whether the nonlocal energy density
then becomes dominant with respect to dust density, depends on the model. The effective equation of state of the universe is approximated by
\be
w_{eff} \approx 4\alpha(\alpha-1)\hat{f}_0a^{-2\alpha}.
\ee
Thus, only if $f_0<0$, do we expect the nonlocal energy density to dominate the matter density. When this happens, our solution 
(\ref{mlimit}) is no longer valid. 
The approximations for the two fields are compared to numerical results in the right panel of figure \ref{appro}. 
When $\alpha<0$, the numerical example for an $f_0>0$ case is depicted in the right panel of figure \ref{eos_no}, and numerical
examples for $f_0<0$ cases are depicted in the right panel of figure \ref{eosn}.

\subsubsection{Power-law model.} Consider the power-law case, $f(\phi) = f_0\phi^n$. For convenience we approximate $\log{\frac{1}{4}ra}+1 \approx \log{ra}$ and
$\sqrt{1+ra} \approx \sqrt{ra}$. Then the solution to the double integral (\ref{di2}) with the lower boundary terms dropped may be written as
\be \label{pot1}
\xi(a) = -f_0\left[\left(2\log{\frac{1}{ra}}\right)^n + (\frac{4}{3})^n  
\frac{n}{(ra)^\frac{3}{2}} \Gamma(n,\frac{3}{2}\log{\frac{1}{ra}})\right].
\ee
From the fact that the incomplete Gamma function behaves asymptotically as $\Gamma(n,x) \approx 
e^{-x}x^{n-1}(1+\mathcal{O}(\frac{1}{x}))$, we see that the second term in Eq.(\ref{pot1}) scales as $\sim (\log{ra})^{n-1}$. 
Therefore the first term, which scales as $\sim (\log{ra})^n$, will eventually dominate. Taking into account the contribution from 
$f(\phi)$ which is equal to $\xi$'s leading contribution, the coupling $\Psi$ reads
\be \label{appro2}
\Psi(a) = 2 f_0\left(2\log{\frac{1}{ra}}\right)^n\left(1 - \frac{n}{6\log{\frac{1}{ra}}}\right).
\ee 
One notes that this satisfies the evolution equation Eq.(\ref{psi_evol}) consistently up to the order $\sim \left(\log{\frac{1}{ra}}\right)^{n-2}$.
Again we have a very simple condition for the nonlocal contribution to grow during an Einstein-de Sitter phase: if $n>0$, the effect 
of the nonlocal terms increases during matter domination, otherwise not. Yet one may check what happens when these terms cannot be compleletely 
neglected. The effective equation of state of our solution is approximated by
\be
w_{eff} \approx (-1)^nf_0\frac{2^{n+1}}{3}n(\log(ra))^{n-1}\left(1+\frac{n-1}{2\log{(ra)}}\right).
\ee
If $(-1)^n f_0 > 0$, we expect that the standard matter domination may be interrupted by the growing nonlocal contribution, but since dust dilutes 
more slowly than the effective nonlocal density, acceleration will not follow. The case $(-1)^n f_0 < 0$ finally represents a scenario where 
$\Omega_m 
\rightarrow 0$ asymptotically, and the nonlocal contribution will eventually dominate the energy density.   
The approximations for the scalar fields are compared to numerical results in the left panel of figure \ref{appro}. 
When $n>0$, the numerical example for a case with $(-1)^{n+1}f_0>0$ is depicted in the left panel of figure \ref{eos_no}, and numerical
examples for the opposite cases are depicted in the left panel of figure \ref{eosn}.

\subsection{Dark energy domination: A sudden singularity and its avoidance}
\label{dari}

Given the conditions above, the universe will enter into an accelerating era. From numerical experimentation, we find that in the
above simple cases this always leads to a so called sudden future singularity \cite{Barrow:2004xh}. This is type II singularity in the 
classification of Ref.\cite{Nojiri:2005sx}. There the second derivative of the scale factor 
diverges at a finite time and when the universe has expanded a finite amount since $a_i$. In addition the first derivative remains finite.
In terms of effective fluid quantities this means that pressure goes to infinity, while the energy density is finite.  

To avoid evolution into such a singularity, several effects have been proposed in frameworks of other models. The backreaction of conformal quantum 
fields might appear near the singularity and change the evolution \cite{Nojiri:2004ip}. Nonperturbative quantum geometric effects can
also resolve the future sudden singularity \cite{Sami:2006wj}. Here we show also that by slight modification of the nonlocal
model by allowing a regulator term $h(R)$ the singularity may cease to exist. 

For definitiveness and simplicity,
let us consider the special case of linear coupling and a linear regulator term, 
\be
f(x) = -\tau x, \quad h(x)=\epsilon x. 
\ee
As discussed in the end of  section (\ref{cosmo}), the action (\ref{h_lag}) allows to integrate out two scalars.
Then $\Psi = -\tau\phi$. The cosmological equations now simplify to \cite{Wetterich:1997bz}
\bea 
3H^2(1-\tau\phi+\frac{1}{2}\epsilon\phi^2) & = & - \frac{1}{4}\dot{\phi}^2 + 3(\tau-\epsilon\phi)H\dot{\phi}, \\ \label{hube1}
-(2\dot{H}+3H^2)(1-\tau\phi+\frac{1}{2}\epsilon\phi^2) & = & (\frac{1}{4}+\epsilon)\dot{\phi}^2 + (\tau-\epsilon\phi)(\ddot{\phi}+2H\dot{\phi}), 
\label{hube2} \\
\ddot{\phi} + 3H\dot{\phi} & = & 6(\tau-\epsilon\phi)(2H^2+\dot{H})^2. \label{hube3}
\eea
The matter sources have been dropped, which is at least near the singularity is a good approximation since there matter becomes
negligible, $\Omega_m \rightarrow 0$. We can write the Friedmann equation (\ref{hube1}) as a first order differential equation,
\be \label{diff}
\frac{da}{a} = \frac{d\phi}{2}\frac{\tau-\epsilon\phi}{1-\tau\phi+\frac{1}{2}\epsilon\phi^2}\left[1+
\sqrt{1-\frac{1}{3}\frac{(1-\tau\phi+\frac{1}{2}\epsilon\phi^2)}{(\tau-\epsilon\phi)^2}}\right].
\ee
This can be integrated. In the case $\epsilon=0$, the solution is
\bea \label{arel}
\frac{a}{a_0} = \sqrt{
\frac{(\tau\phi_0-1)(1+y)(1-y_0)}{(\tau\phi-1)(1+y_0)(1-y)}}e^{y_0-y}, \\ \nonumber y \equiv 1+\frac{1}{3\tau^2}(\tau\phi-1).
\eea
Because of difficulties in inverting this relation, we proceed by considering cosmology as a function of the scalar field $\phi$. From Eq.(\ref{diff}), we 
can relate the Hubble rate to the time derivative of the field for each value of the scalar. Then, we can relate effective density $\rho_\phi = 3H^2$ 
to the square of the field 
derivative, given the value of the field. To find a similar expression for the effective pressure $p_{eff} =  -(2\dot{H}+3H^2)$, we must first use equation 
(\ref{hube2}) and solve $\ddot{\phi}$ by using (\ref{hube3}). The full result of this algebra, taking into account a nonzero $\epsilon$, is
\be \label{pres}
w_\phi = \frac{-6(1+\tau(\phi+5\tau)) + c(\tau,\phi,\epsilon)\epsilon
- 4\sqrt{6}(1-6\epsilon)(\tau-\epsilon\phi)^2
\sqrt{6+\frac{2\phi}{\tau-\epsilon\tau} + 
\frac{-2+\epsilon\tau}{(t-\epsilon\tau)^2}}
}{3(-2 + 2\tau(\phi + 3\tau) - \epsilon\phi(\phi + 12\tau) + 6\epsilon^2\phi^2)}, 
\ee
where
\be
c(\tau,\phi,\epsilon) = -3(-8+(1+2\epsilon(-7+24\epsilon))\phi^2+4(7-24\epsilon\tau\phi + 48\tau^2)).
\ee
Consider first the case that $\epsilon=0$. We then see that the equation of state diverges at finite $\phi_{crit} = 3\tau-\frac{1}{\tau}$. 
The relation (\ref{arel}) shows that this happens at a finite value of the scale factor, and the Friedmann equation (\ref{hube1}) shows that the 
density is also finite at $\phi = \phi_{crit}$. Thus we have a future sudden singularity.

We then restore the regulator term $\epsilon>0$ into the operator to show that then the singularity may disappear. Now, 
the denominator of (\ref{pres}) vanishes if
\be \label{divp}
\phi_{crit} = \frac{\tau-6\epsilon\tau \pm 
\sqrt{(6\epsilon-1)(2\epsilon-\tau^2)}}{\epsilon-6\epsilon^2}
\ee
One may adjust $\epsilon$ in such a way that the denominator in (\ref{pres}) is always 
nonzero, i.e. that the equation for the critical $\phi_{crit}$ as given by equation (\ref{divp}) does not have real solutions. 
To achieve this when
the absolute value of $\tau$ exceeds $|\tau|>1/\sqrt{3}$, one has to set $1/6 \le \epsilon 
\le \tau^2/2$. When the absolute value of $\tau$ is smaller, $|\tau|>1/\sqrt{3}$, one has to set
$\tau^2/2 \le \epsilon \le 1/6$. Thus we have shown that there are values of $\epsilon$ which completely 
remove the singularity.

\subsection{Reconstruction of cosmology}
\label{reco}

One might approach the dynamical problem from the other way and ask what kind of modifications $f$ would be needed to generate a given expansion history.
Previously such reconstructions have been considered for scalar-tensor theories \cite{Boisseau:2000pr}, vector-tensor theories 
\cite{Koivisto:2008ig,Koivisto:2008xf}, coupled scalar theories \cite{Tsujikawa:2005ju}, $f(R)$ theories \cite{Capozziello:2005ku}
and the Gauss-bonnet theory \cite{Nojiri:2006je}; for reviews, see \cite{Sahni:2006pa, Nojiri:2006be}.

Since our action contains an arbitrary function $f$, one might think that by adjusting the function suitably one could model an arbitrary background
expansion. The recipe for finding the function $f$ can be given quite simply. A given expansion history $a(t)$ is equivalent to a given function 
$w_{eff}(a)$. Therefore, the variable $X(a)$ and thus the field $\phi(a)$ can be found by integrating the equation of motion (\ref{x_evol})
\be \label{x_evol2}
X^* =  \frac{3}{2}\left[(w_{eff}-1)X  + 3w_{eff}-1\right].
\ee
It is possible to derive a similar equation for the other field $\Psi$ as well,
\be \label{psi_evol2}
\Psi^{**} + \frac{1}{2}(7-3w_{eff})\Psi^* + 3(1-w_{eff})\Psi = -3(1-w_{eff}) + 3\Omega_m(1-w_m). 
\ee
Therefore, a given background expansion indeed determines uniquely the form of both the fields $\phi$ and $\Psi$. One can write the Friedmann equation 
in the form 
\be \label{friedmann_2}
1+\Psi = \Omega_m + \frac{1}{3}(2f'(\phi)X - \Psi^*)X - \Psi^*.
\ee
It is then straightforward to insert the solutions for $\phi$ from Eq.(\ref{x_evol2}) and the solutions for $\Psi$ from Eq.(\ref{psi_evol2}) into 
the Friedmann
Eq.(\ref{friedmann_2}) to find $f'(\phi)$ as a function of the scale factor $a$. The coupling as a function of the scale factor is then given by the 
integral
\be \label{coupling}
f(a) = \int_{a_i}^a 2aX(a)f'(\phi) da.
\ee
If one is then able to invert $\phi(a)$ to get $a(\phi)$, the functional form of the coupling is finally found as $f(\phi) = f(a(\phi))$.
We demonstrate this procedure with an explicit example below.

Consider cosmological expansion with the constant effective equation of state $w_{eff}=w$. We can then easily solve $\phi$ from Eq.(\ref{x_evol2}),
\be \label{w_phi}
\phi(a) = \frac{2(1-3w)}{w-1}\log{a} + \phi_1a^{\frac{3}{2}(w-1)} + \phi_0,
\ee
and check that the result also agrees with the integral (\ref{di}), $\phi_1$ and $\phi_0$ being the integration constants. In equation 
$(\ref{psi_evol2})$ we set now $\Omega_m(1-w_m) = \Omega^0_m a^{3(w-w_m)}(1-w_m)$. The equation is then solved by
\be \label{w_psi}
\Psi(a) = -1 + \left(\frac{w_m-1}{w-1}\right)\Omega_m^0a^{3(w-w_m)} + 
\frac{\Psi_1}{a^{2}} + \frac{\Psi_2}{a^{\frac{3}{2}(w-1)}},
\ee
where the two first terms give the special solution, simply a constant, and the two last terms give the general solution to the homogeneous 
equation with the 
constants $\Psi_1$ and $\Psi_2$ which should be interpreted as the boundary conditions. 
To get a nice expression for $f'(\phi)$ from the Friedmann
equation (\ref{friedmann_2}), we approximate $\phi(a)$ by the constant and the logarithmic term in (\ref{w_phi}). This is justified when
$w<1$, so that the second term decays fastest, and when $a$ is large enough. Similarly, if we assume again large enough $a$, and 
the effective expansion is nonphantom when the physical matter content doesn't decay faster than dust, or more generally 
$w > 2w_m-1$, then the solution (\ref{w_psi}) may be approximated by setting $\Psi_2=0$. Then we obtain
\be
f'(\phi(a)) = 
\frac{3(w-w_m)(3-w-2w_m)\Omega_m^0 a^{2+3(w-w_m)} + (w-1)(1+3w)\Psi_1}{2(1-3w)^2a^2} 
\ee
and by integrating furthermore 
\be
f(\phi(a)) = 
\frac{-2(3-w-2w_m)\Omega_m^0 a^{2+3(w-w_m)} + (w-1)(1+3w)\Psi_1}{4\left(1+w(3w-4)\right)a^2}, 
\ee
where the integration constant should be dropped as it corresponds to a constant shift of $\kappa^2$. By inverting the relation Eq.(\ref{w_phi}) we 
finally get the function as
\be
f(\phi) = \frac{1}{4\left(1+w(3w-4)\right)}
\left[ (w-1)(3w+1)\Psi_1e^{\frac{(1-w)}{(1-3w)}(\phi-\phi_0)}
+
2(w+2w_m-3)\Omega^0_me^{\frac{3}{2}\frac{(1-w)(w_m-w)}{(1-3w)}(\phi-\phi_0)}\right].
\ee
The main result about this example is that expansion with a constant equation of state (with or without matter) corresponds to an exponential 
coupling. Details are discussed below. 

We find two possibilities: a vacuum solution and a scaling solution.
\begin{itemize}
\item
In the case that $w < w_m - 2/3$ the asymptotic form of the matter coupling becomes tdominated by the first term inside the square 
brackets at large times. Thus, if we assume the effective equation of state to be more negative the $\Psi_1$-term dominates at large $a$ instead of 
the matter term. Since the matter has diluted away, this corresponds to the vacuum solutions found by Jhingan {\it et al} \cite{Jhingan:2008ym}.
As a function of $\alpha$, the expansion rate of the vacuum corresponds to 
\be
w_{eff} = \frac{\alpha-1}{3\alpha-1}.
\ee
With the addition of dust, these solutions become unstable. Thus one cannot reach them from the matter-dominated era.   
\item
If either the boundary condition is exactly $\Psi_1=0$, or the expansion is less accelerating by
$w > w_m - 2/3$, then we find that the second term inside the square brackets dominates asymptotically and the solution is 
determined solely by the matter content. As one can deduce from Eq.(\ref{w_psi}), in this solution the ratio 
$\Omega_m/\Psi$ is a constant. In fact this is a scaling solution, since effective fractional matter density $\Omega_m/(1+\Psi)$ 
is asymptotically a constant, since $\Psi$ grows quickly to large values. This corresponds to a huge gravitational constant,   
which can be immediately excluded on observational grounds. The effective equation of state of this scaling solution is given
as a function of $\alpha$ by
\be \label{scaling}
w_{eff} = \frac{1}{2}(1+w_m) - \alpha - \frac{1}{6}\sqrt{24\alpha - 36w_m + (3-6\alpha+3w_m)^2}.
\ee
Since this is positive definite for $w_m=0$ and $\alpha<0$, one 
cannot use 
it to obtain accelerating from the dust dominated
era.
\end{itemize}




\begin{table}
\begin{tabular}{|c|c|c|c|c|}
\hline
$n$  & $f_0$ & $w_{eff}$ & $\phi$ & $\Psi$ \\
\hline
$1$ & $1.852216e-02$ & $-9.701123e-02$ & $-1.705136e+01$ & $-6.316559e-01$ \\ \hline
$2$ & $-1.068728e-03$ & $-2.366542e-01$ & $-1.673209e+01$ & $-5.597827e-01$ \\ \hline
$3$ & $6.090079e-05$ & $-4.194428e-01$ & $-1.656208e+01$ & $-4.989559e-01$\\ \hline
$4$ & $-3.487378e-06$ & $-6.412859e-01$ & $-1.643555e+01$ & $-4.461259e-01$\\ \hline
$5$ & $2.013315e-07$ & $-8.990863e-01$ & $-1.633576e+01$ & $-4.010074e-01$\\ \hline
$6$ & $-1.172498e-08$ & $-1.196474e+00$ & $-1.625593e+01$ & $-3.629566e-01$\\ \hline
$7$ & $6.875924e-10$ & $-1.517736e+00$ & $-1.618961e+01$ & $-3.302487e-01$\\ \hline
$8$ & $-4.063656e-11$ & $-1.891849e+00$ & $-1.613562e+01$ & $-3.027929e-01$ \\ \hline
$9$ & $2.412112e-12$ & $-2.252064e+00$ & $-1.608829e+01$ & $-2.783980e-01$\\ \hline
$10$ & $-1.440718e-13$ & $-2.665776e+00$ & $-1.604892e+01$ & $-2.576894e-01$\\ \hline
\end{tabular}
\caption{\label{tab} Numerical values for the power-law model with different exponents $n$. We report the scale $f_0$ which yields
$\Omega^0_m = 0.3$. For even $n$ this number must be negative. The following three columns then give the effective equation of the state and the values 
of the two fields evaluated today.  
}
\end{table}

\section{Conclusions}
\label{conclu}

We studied the dynamical evolution of cosmological expansion in a class of nonlocal gravity models extending general relativity
by the addition of terms of the type $f(R/\Box)$, which are motivated by quantum corrections to the gravity action. 
The corrected action was rewritten as a multi-scalar-tensor theory, where several aspects of the theory such as 
its physical degrees of freedom, become more transparent. We also considered Weyl rescaling, but since the Einstein frame 
of these theories was not practical we performed our main analysis in the physical Jordan frame. The main result of this 
paper is the first explicit demonstration that these models can explain the present acceleration of the universe with 
simple forms of $f(R/\Box)$ involving only Planck scale constants.

In particular, power-law models and exponential models were used as examples. Schematically, we may summarize the different possibilities of 
cosmological evolution in the exponential model as follows:
\begin{displaymath}
\xymatrix{
f=f_0e^{\alpha\phi} \ar[r]^{\alpha<0} \ar[d]_{\alpha>0} & \text{Nonlocal effect} \ar[d]_{f_0>0} \ar[dr]^{f_0<0} &         & \\
\text{Matter domination}                                & \text{Scaling, Eq.(\ref{scaling})} & \text{Acceleration} \ar[r] & \text{Singularity}
}
\end{displaymath}
Note that in the case $\alpha>0$ one may have effects in the early universe. 
The power-law model features analogous possibilities. Now however the scaling is not exact, but the significance of the nonlocal 
correction decays and leads back to the usual Einstein - de Sitter stage. Schematically, we summarize this as: 
\begin{displaymath}
\xymatrix{
f=f_0\phi^n \ar[r]^{n>0} \ar[d]_{n<0} & \text{Nonlocal effect} \ar[d]_{(-1)^{n}f_0>0} \ar[dr]^{(-1)^{n}f_0<0} &     & \\
\text{Matter domination}                                & \text{Slows down expansion} \ar[l] & \text{Acceleration} \ar[r] & \text{Singularity}
}
\end{displaymath}
Again, though this was not in the focus of the present study, $n<0$ may lead to a decaying effect which, with large enough scale $f_0$, can influence
the evolution of the early universe.  
Let us note that in both of the above cases, the conditions for acceleration to occur at late times imply $f'(\phi)>0$. According to 
Eq.(\ref{set_phi}), this is precisely the condition for the field $\phi$ not to be a ghost. The consistency condition for the other field, 
$\Psi>-1$, ensuring the  positivity of the graviton energy, is satisfied by all of the models we have considered. 

We sketched also a reconstruction method and applied it to find scaling solutions including matter and vacuum solutions with constant $w_{eff}$.
These vacuum solutions have been found previously and shown to feature a possible de Sitter expansion \cite{Nojiri:2007uq} or effective dark 
energy with a range of equations of state extending to phantom regime \cite{Jhingan:2008ym}. We confirmed and generalized these solutions, and 
noted that by addition of matter they become unstable, and thus cannot be reached in realistic cosmology. This shows that a detailed analysis of 
dynamics, without neglecting the effect of matter, is necessary in assessment of the cosmological relevance of these models. One could draw some 
analogy with previous findings about the cosmological evolution in the Gauss-Bonnet dark energy 
\cite{Nojiri:2005vv}, or in the gravity models of the $f(R)$ type \cite{Carroll:2003wy}: though standard cosmological solutions 
could be (approximately) recovered at higher curvature and accelerating solutions would exist in vacuum at low curvature, the conditions for 
the cosmological dynamics to connect the first with the other may be very nontrivial \cite{Koivisto:2006xf,Amendola:2006we}. To address the 
cosmological viability, it is also crucial to 
investigate the perturbations, as in the two classes of models mentioned above \cite{Koivisto:2005yc,Koivisto:2006ai}; for the 
present models this is left for future study. The form of the modifications is such that the impact to local gravity
experiments could be expected to be negligible, but this should be checked in detail as well.

A rather dramatic prediction of these models is that the present acceleration is a manifestation of our universe plunging into
a sudden singularity. 
Though it awaits in the future and thus does not affect our present observations, its presence
may be considered as an unpleasant property of the models signalling the need to choose more carefully the phenomenological
gravity action. 
In our tentative example, we showed explicitly the disappearance of the singularity by shifting $\Box \rightarrow \Box + 
\epsilon R$. Finally, it is worth mentioning that the particular form of nonlocalities studied here is a direct 
generalization of an action previously considered in the quite different context of stabilizing the 
Euclidean gravity \cite{Wetterich:1997bz}. It is interesting that one may constrain nonperturbative quantum corrections to gravity by their 
cosmological effects at the present day, and that nonlocality may provide the required new physics to such seemingly detached major problems as the 
black hole information paradox and the fine-tuning of the cosmological constant. 


\bibliography{nl}

\end{document}